# Dynamic lens and monovision 3D displays to improve viewer comfort

**Paul V. Johnson[1], Jared AQ. Parnell[2], Joohwan Kim[3], Christopher D. Saunter[2], Gordon D. Love[2,*] and Martin S. Banks[1,3,†]**

[1]*UC Berkeley – UCSF Graduate Program in Bioengineering, Berkeley, CA 94720, USA*
[2] *Durham University, Dept. of Physics, Durham DH1 3LE, UK*
[3]*Vision Science Program, School of Optometry, University of California, Berkeley, CA 94720, USA*

[*]g.d.love@durham.ac.uk, [†]martybanks@berkeley.edu

**Abstract:** Stereoscopic 3D (S3D) displays provide an additional sense of depth compared to non-stereoscopic displays by sending slightly different images to the two eyes. But conventional S3D displays do not reproduce all natural depth cues. In particular, focus cues are incorrect causing mismatches between accommodation and vergence: The eyes must accommodate to the display screen to create sharp retinal images even when binocular disparity drives the eyes to converge to other distances. This mismatch causes visual discomfort and reduces visual performance. We propose and assess two new techniques that are designed to reduce the vergence-accommodation conflict and thereby decrease discomfort and increase visual performance. These techniques are much simpler to implement than previous conflict-reducing techniques. The first proposed technique uses variable-focus lenses between the display and the viewer's eyes. The power of the lenses is yoked to the expected vergence distance thereby reducing the mismatch between vergence and accommodation. The second proposed technique uses a fixed lens in front of one eye and relies on the binocularly fused percept being determined by one eye and then the other, depending on simulated distance. We conducted performance tests and discomfort assessments with both techniques and compared the results to those of a conventional S3D display. The first proposed technique, but not the second, yielded clear improvements in performance and reductions in discomfort. This dynamic-lens technique therefore offers an easily implemented technique for reducing the vergence-accommodation conflict and thereby improving viewer experience.

# 1. Introduction

When a viewer looks at an object in the natural environment, the two eyes must be directed to that object. Without appropriate vergence eye movements to align the lines of sight, double vision would occur. At the same time, the eyes must accommodate on the fixated object so that the retinal images of the object are sharp. If the eyes did not accommodate accurately, blurred vision would occur. When the viewer looks from one object to another, vergence and accommodation must change accordingly. Because the distances to which the eyes must converge and accommodate are almost always the same in the natural environment, vergence and accommodative responses are coupled neurally. As a consequence, changes in accommodation evoke changes in vergence and changes in vergence evoke changes in accommodation. A benefit of the coupling is that vergence and accommodative response happen more quickly when they occur together. Specifically, vergence and accommodation are faster when disparity and blur specify the same change in distance as opposed to when they specify different changes in distance [1-5]. The left panel of Fig. 1 illustrates how vergence and accommodation change together in natural viewing.

Stereoscopic 3D (S3D) displays deliver slightly different images to the left and right eyes in order to create binocular disparity and thereby produce an enhanced impression of depth. Again the eyes must make vergence eye movements to different distances in the simulated scene: converging for near objects (crossed disparity) and diverging for far ones (uncrossed disparity). The eyes must also accommodate, but now to the distance of the screen rather than the distance of the simulated object. The mismatch between vergence distance and accommodation distance disrupts the normal vergence-accommodation coupling. The second column in Fig. 1 schematizes this mismatch. The vergence-accommodation conflict that occurs with conventional S3D displays causes some, perhaps all, of the visual discomfort (eye fatigue, eye irritation, blurry vision, headache, nausea, etc.) that accompanies prolonged viewing of such displays [6-12].

The vergence-accommodation conflict occurs in all S3D displays currently on the market, regardless of the method used to deliver the appropriate content to each eye (e.g., temporal interlacing, spatial interlacing, color anaglyph). If the conflict is large, the stimulus is likely to appear blurred, double, or both [13]. It is therefore critically important to know the ranges of vergence and accommodation distances that can be presented without undesirable side effects. Shibata and colleagues measured a "zone of comfort," or a range of vergence and accommodation distances that does not cause discomfort [12]. One can apply knowledge of this range to limit the range of disparities presented and thereby minimize discomfort. But this limits that range of possible perceived depths and therefore does not allow the presentation of dramatic depth effects.

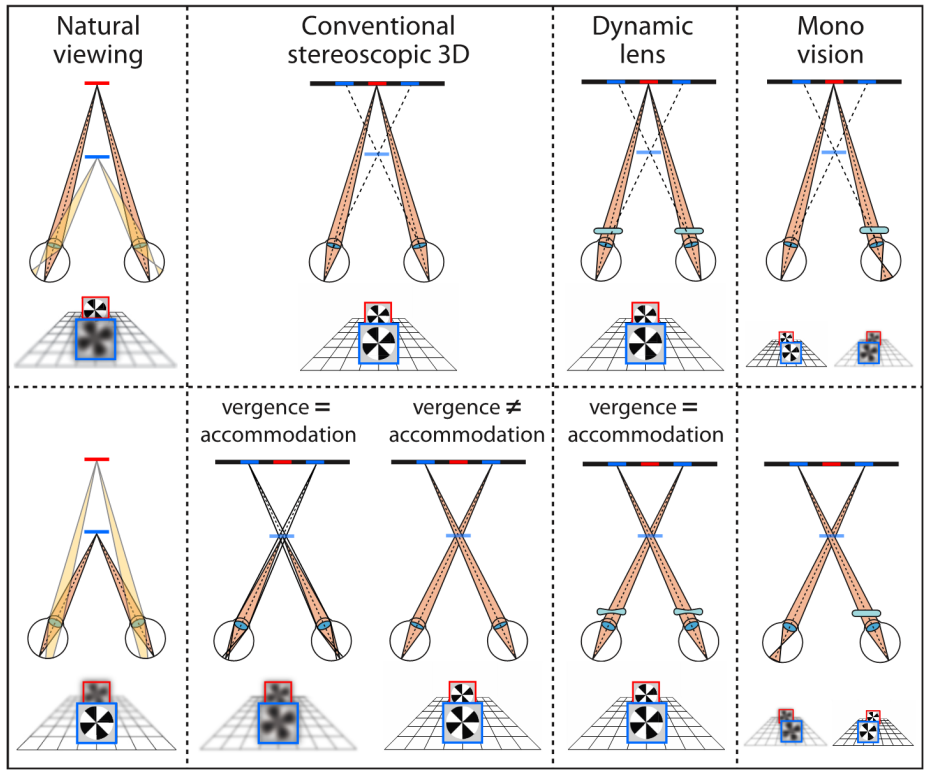

Fig. 1. Focus cues in different viewing conditions. In each panel, the upper part is an overhead view of the situation and the lower part is a schematic of what a viewer would see. The upper row shows these when the viewer fixates the far (red) object and the lower rows shows them when the viewer fixates the near (blue) object. In natural viewing (left column), vergence and accommodation distances are the same. In conventional stereoscopic 3D displays (second column), vergence distance varies with the disparity of the simulated object while accommodation distance is fixed at the display screen. Thus, the two distances are often dissimilar. In the proposed dynamic-lens display (third column), accommodation distance can be adjusted depending on the content being displayed. This is implemented by changing the power of a lens in front of each eye. In the proposed monovision display (right column), fixed lenses of different powers are placed in front of the two eyes. Here the power of the lens in front of the right eye is lower than the the power of the lens in front of the left eye. Thus, there are two accommodative distances, one for the left eye and one for the right.

Because of the desire to retain dramatic 3D effects while reducing visual discomfort, there have been many attempts to construct S3D displays that reproduce focus cues and thereby decrease vergence-accommodation conflicts. They can be divided into three categories: volumetric, multi-plane, and light-field displays.

Volumetric displays place light sources (voxels) in a 3D volume by using rotating display screens [14] or stacks of switchable diffusers [15]. These allow correct vergence and accommodation cues, but the scene is restricted to the size of the display volume, and the large number of required addressable voxels places practical limits on resolution. An additional serious limitation is that these displays present additive light, creating a scene of glowing, transparent voxels. This makes it impossible to reproduce occlusions and specular reflections correctly.

Multi-plane displays are a variation of volumetric displays in which the viewpoint is fixed. Such displays can in principle provide correct depth cues, including focus cues, with conventional display hardware. In multi-plane displays, images are drawn on presentation planes at several focal distances for each eye, enabling both vergence and accommodation cues. Such displays have been made using a set of beam splitters [16,17] and by time-multiplexing with high-speed switchable lenses [18,19] to superimpose multiple presentation planes additively on the viewer's retinas. Current implementations support high-resolution imagery by using the full resolution of a conventional display. Focus cues are correct for simulated objects lying on one of the presentation planes. By using a depth-weighted blending rule to assign intensity to pixels on surrounding planes, focus cues are approximately correct for simulated objects positioned in between planes [16,20]. Head-mounted versions of multi-plane displays have been developed [21,22]. The most serious limitation of the multi-plane approach is that it requires very accurate alignment between the viewer's eyes and the presentation planes. Thus, the positioning between the display and viewer's eyes must be precise and stable, which limits the practical utility of the displays.

The third category is light-field displays that are designed to reproduce a four-dimensional light field, allowing glasses-free viewing with stereoscopic and parallax cues. Initial approaches used lenticular arrays [23,24] and parallax barriers [25,26] to direct exiting light along different paths. Later developments explored compressive techniques based on multi-layer architectures [27-30]. In principle, a light-field display can produce accurate focus cues because a light field theoretically encodes the full radiance distribution emitted from the scene. However, for normal viewing distances, presenting focus cues to human viewers requires a display with extremely high angular resolution [31-33]. Maimone and colleagues [34] proposed an architecture that uses a combination of a light-attenuating liquid-crystal stack and a high-resolution backlight to steer light in the direction of the viewer, potentially supporting accommodation. Currently, resolution requirements and computational workload are too demanding to make a practical light-field display that supports focus cues.

None of these displays have been widely used because of one or more of the following drawbacks: inability to support occlusions and reflections (but see [35]), small field of view, large physical size, limited number of focal states, requirement for custom hardware or imaging optics, loss of spatial resolution, etc. For example, consider multi-plane devices, and in particular the device we described previously [18]. In this display, a switchable lens rapidly changes power as different depths are displayed on the screen. The display creates a convincing 3D volume and greatly reduces vergence-accommodation conflict, but it has the major disadvantage that head position must be known and fixed for parallax at the retinas to be correct.

We propose two display techniques that involve placing lenses between the viewer's eyes and the display screen. The first, which we call the *dynamic-lens technique*, uses lenses that can change focal length over time. This is illustrated in the third column in Fig. 1. If the lens powers are changed in synchrony with the distance of fixated objects in the otherwise conventional stereoscopic content, the match between vergence and accommodation distances is restored and the vergence-accommodation conflict is minimized. This requires a reasonably accurate estimate of the viewer's fixation distance, a point we discuss in detail later. The second proposed technique, which we call the *monovision technique*, is even simpler. It is illustrated in the fourth column in Fig. 1. We place a fixed lens in front of one eye and present otherwise conventional stereoscopic content. Depending on the accommodative state of the viewer's eyes, the retinal image in one eye will be in better focus than in the other eye. A change in accommodative state can cause that relationship to reverse. This is very similar to monovision, a clinical treatment for presbyopia (age-related reduction in accommodative range). In this treatment, the optical correction for one eye is appropriate for distance viewing while the correction for the other eye is appropriate for reading distance [36]. In adapting this approach to stereoscopic displays, we hypothesized that the viewer's binocular percept will be dictated by the eye whose focal distance is closer to the vergence distance of the fixated object, and therefore that the vergence and accommodative responses will be more similar than they would be in a conventional S3D display. Fig. 2 demonstrates this: It is a stereoscopic photograph with the camera's focal distance set differently for the two eyes.

When one cross-fuses the photograph (directing the left eye to the right image and the right eye to the left image), the binocularly fused image appears generally sharp. This means that the right eye dictates the percept where the left-eye's image is blurred and vice versa. We hoped that the monovision technique would reduce the vergence-accommodation conflict leading to a reduction in visual discomfort.

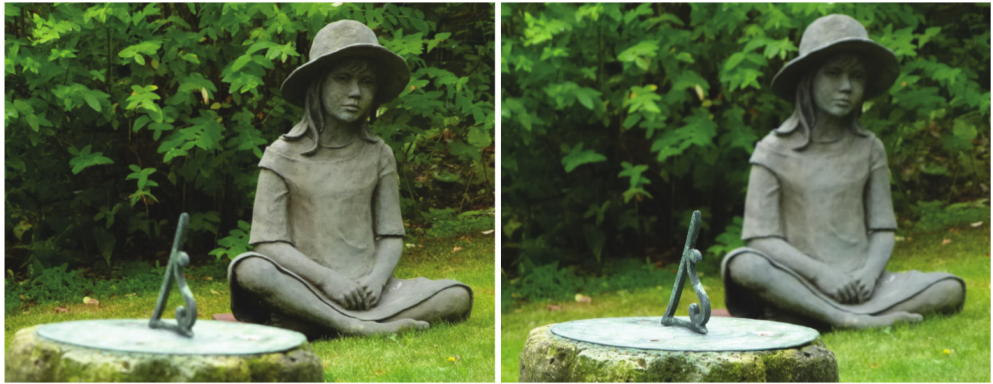

Fig. 2. Stereoscopic photograph demonstrating the principle behind the monovision technique. The left image was taken with the camera focused at the statue of the girl. The right image was taken with it focused at the sun dial. Cross-fuse the photograph: i.e., direct the left eye to the right image and the right eye to the left image. The binocularly fused image in the middle contains contributions from both eyes. Notice that the fused image appears generally sharp: sharper than either of the monocular images. This shows that the right eye dictates the binocular appearance in regions where the left-eye image is blurred and that the left eye dictates the appearance in regions where the right-eye image is blurred.

We implemented the two proposed techniques and assessed their efficacy relative to a conventional S3D display by measuring visual performance and visual discomfort.

## 3. Experimental Details: Displays and Optics

### 3.1 Dynamic-lens system

The dynamic-lens system is the same as a conventional stereoscopic display system except that there is a variable-power lens between each eye and the display screen. The power of the lenses is adaptively adjusted so that the distance to which the eyes must accommodative to see sharp images is the same as the distance to which the eyes must converge to see a single, fused image. In this way, the vergence-accommodation conflict is eliminated, or at least greatly reduced, in comparison to a conventional stereoscopic display.

The display itself is a commercially available S3D display (23" LG Cinema 3D D2342) that uses spatial interlacing to send the left- and right-eye content to the appropriate eyes. Viewing distance was 1.77m; at that distance pixels subtended 0.52 arcmin. Other types of displays (e.g., temporal interlacing, color anaglyph) could also be used to create the dynamic-lens system.

The lenses are Optotune lenses (Optotune, Dietikon, Switzerland, EL-10-30-VIS-LD) that can be driven to different focal lengths dynamically. The lenses are connected to a driver (Optotune USB Lens Driver 4, OEM version) hosted by a Mac Pro desktop computer (Apple Inc., Cupertino, CA). The focal length of each lens can be adjusted to a specific value within milliseconds by supplying current in the range of 0-300mA. The focal length changes are instantiated by changes in lens curvature. The driving current was adjusted using a Python API provided by Optotune. Screen distance was 1.77m (0.57D). With the range of focal length changes, the resulting range of accommodative distances was 0.48–3.2m (2.06–0.31D).

The Optotune lens can enable larger ranges, but we did not require them. We calibrated the lenses to ensure an accurate mapping between current value and focal length. For current values of 175, 200, and 225mA, we manually focused a camera looking through the lens on a high-resolution Siemens star (a spoked-wheel pattern) that was printed and placed on the display screen. We adjusted focus until we obtained the sharpest image of the star. We then moved the camera, without changing its focus, to view a Siemens star without the lens. We found the distance of the star that yielded the sharpest image. We repeated the procedure for each of the three current values. We fit a plot of current vs focal distance with a line and used this to determine the current values needed during the course of the experiments.

The third column of Fig. 1 illustrates how the lenses in principle can drive accommodative distance to match vergence distance. The discomfort associated with the vergence-accommodation conflict is presumably the result of mismatches in vergence and accommodative *responses* (i.e., the eyes being converged on a stimulus nearer than the display screen while the eyes are accommodated to the distance of the screen). In attempting to reduce the conflict in responses to zero, one would logically match the distances of the vergence and accommodative *stimuli*. But the distance of the vergence stimulus depends on what the viewer is looking at. Thus, the dynamic-lens system requires a reasonably accurate estimate of the distance of the fixated stimulus. We avoided this issue in the work reported here by providing a fixation target and instructing subjects to keep fixating that target as it changed distance. We discuss later how one could implement the system with gaze prediction or gaze measurement. Thus, the system in the experiments always had knowledge of the current fixation distance. By reducing the vergence-accommodation conflict, the dynamic-lens display system should provide a more comfortable viewing experience than conventional S3D displays. Moreover, vergence and accommodative responses should occur more quickly than in conventional displays because the normal coupling between vergence and accommodation would drive the responses to the same distance. This should increase the ability to binocularly fuse stimuli quickly and thereby improve visual performance.

*3.2 Monovision system*

Monovision is a fairly common optometric/ophthalmic method for treating presbyopia (age-related loss of the ability to accommodate). In monovision, one eye is given an optical correction appropriate for far distance and the other eye a correction appropriate for near. Typically, the difference in optical power in front of the two eyes is 1–1.5D [36]. The idea is that the patient will use the eye corrected for distance when looking at far objects and will switch to the eye corrected for near when doing nearwork. Behind this idea is the assumption that the binocular visual system will suppress the image from the blurrier eye so that the fused percept will be reasonably sharp enabling the patient to see relatively clearly at both distance and near. Fig. 2 demonstrates this phenomenon. 60–75% of patients can manage the differential correction to the two eyes, but that percentage decreases significantly as the difference in optical power increases [37]. Accommodation is yoked in humans meaning that when one eye changes accommodative state by a certain amount, the other eye changes by the same amount [38-39]. If a person is pre-presbyopic (i.e., they can still accommodate), accommodative responses are determined by the sighting eye [40]. Not surprisingly, the precision of stereopsis (e.g., stereoacuity) is worse with monovision compared to full correction of both eyes [41]. We adopted this clinical procedure for use with a stereoscopic display.

In the monovision system, we placed a fixed lens in front of one of the subject's eyes. The display screen was very similar (23" LG 3D D2343P) to the one in the dynamic-lens system and the means of presenting separate left- and right-eye images was the same. The setup is schematized in the right column of Fig. 1. The viewing distance to the display was 2m (0.5D) and each pixel subtended 0.46arcmin.

Our intention in constructing this system is that the cyclopean percept and accommodative response would be based on the eye corrected for distance when viewing an object at a long simulated distance and that the percept and response would be based on the eye corrected for near when viewing a near simulated object. In this way, the cyclopean percept should remain relatively sharp and accommodative response should be similar to vergence response thereby minimizing the vergence-accommodation conflict.

We ran two conditions: a conventional condition in which the two eyes had the same optical correction and a monovision condition in which the eyes had different corrections. To implement the conditions, we used two pairs of spectacles. The first had zero power in both eyes and the second had –1D in the right eye and 0D in the other. The choice of –1D is a tradeoff between having too small an offset (thereby creating a limited workspace) and having too large an offset (thereby increasing the number of subjects who experience discomfort due to the inter-ocular difference in focus [37]).

## 3. Visual Discomfort Measurements

### 3.1 Experimental Details

We conducted two experiments—one with the dynamic-lens system and one with the monovision system—to determine whether less visual discomfort is experienced with these systems relative to the discomfort associated with conventional stereoscopic displays. In both experiments, we presented a stimulus like the one in Fig. 3. A white diamond on a gray background moved back and forth in depth from +1.5D (in front of the screen) to –0.25D (behind the screen). It took 5.5sec to travel from one extreme to another, pausing at each end for 0.5sec. The diopter values are the distances of the diamond to the viewer relative to the distance from the viewer to the display. One of the circles also contained a small positive (crossed) disparity, while the other three circles had zero disparity relative to the diamond. As the diamond appeared to come forward from the screen and to recede to behind the screen, one of the circles would appear to come forward for 1sec relative to the diamond. The relative disparity of the target circle varied between 0–4 arcmin. Subjects indicated which of the four circles appeared nearer in depth than the other three: a 4-alternative, forced-choice task. If a subject did not respond within 4sec after the target circle was extinguished, the computer assigned a random response, yielding on average 25% correct performance for such trials. Feedback about the correctness of each response was provided. The positions of the circles within the diamond were randomly perturbed so the task could not be done from one eye's image alone. Stimuli were generated using Python's PsychoPy library. The psychometric data were fit with a cumulative Gaussian function using a maximum-likelihood criterion [42-44]. We define the threshold disparity as the value at which the fitted Gaussian crossed 62.5%. When we averaged data across subjects, we did so by pooling the psychometric data from all subjects and then fitting the pooled data with a cumulative Gaussian function.

There were two experimental conditions: one in which the systems were activated (i.e., the focal length of the lens matched the stimulus for the dynamic-lens experiment or the -1D lens was used in the monovision experiment), and a condition that mimicked a conventional stereoscopic display (i.e., focal length was fixed in the dynamic-lens experiment; both eyes had 0D in front of them in the monovision experiment). The dynamic-lens and monovision experiments both consisted of two sessions, one for each experimental condition presented in random order. Sessions lasted ~10min each. There was a mandatory break of at least 15min between sessions. After each session, subjects filled out a symptom questionnaire asking them to rate, on a scale of 0-6, how they felt in terms of eye tiredness, blurry vision, nausea, neck and back tiredness, eye strain, and headache. At the end of the experiment, subjects filled out a comparison questionnaire that asked which session they preferred in terms of general fatigue, eye irritation, headache, nausea, and overall preference. The whole experiment lasted ~1 hour including training and debriefing.

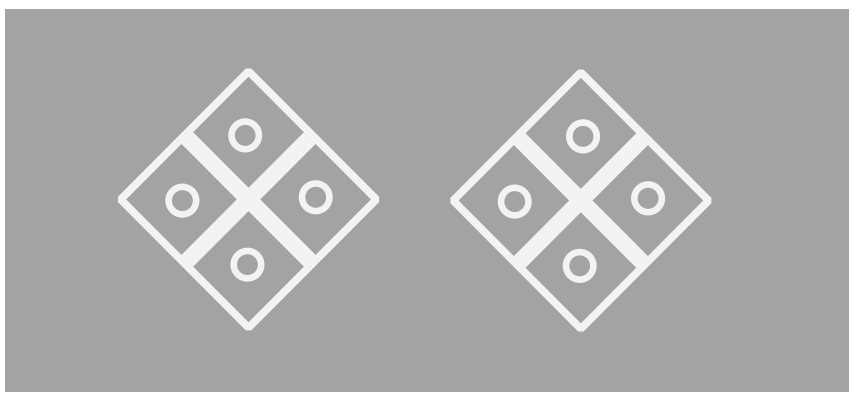

Fig. 3. Stimulus used in the visual discomfort experiments..Cross fuse to see the binocular stimulus. Subjects indicated which of the circles was popping out in depth (a 4-alternative forced-choice task). In this case, the top circle is the correct response. The diamond and X structure moved back and forth in depth, and the circles periodically appeared for 1sec followed by a 2sec break in which the viewer could respond.

Subjects varied in age from 18–30 years. All had normal or corrected-to-normal vision and good accommodative ranges. The stereoacuity of subjects in the dynamic-lens experiment was tested prior to starting the experiment using the Stereo Fly Test (Western Ophthalmics Co., Lynnwood, WA, USA), a standard optometric test. The binocular function of subjects in the monovision experiment was also tested by determining whether they were able to cross-fuse stereo images. 23 subjects participated in the dynamic-lens experiment and 18 in the monovision experiment. None of the subjects were aware of the experimental hypotheses. Appropriate consent and debriefing were done according to the Declarations of Helsinki. We excluded subjects from further analysis if they were unable to reliably indicate the target circle. Specifically, we excluded subjects who could not do better than 60% correct at the largest disparity of 4 arcmin. This criterion led to the exclusion of seven subjects in each of the two experiments.

*3.2 Visual discomfort results for the dynamic-lens system*

The disparity-detection data allowed us to determine whether the magnitude of the vergence-accommodation conflict affected the ability to detect small disparities. Fig. 4 shows the results for one typical subject in the dynamic-lens experiment. Threshold was lower in the dynamic-lens condition than in the fixed-lens condition, indicating that minimizing the vergence-accommodation conflict allowed this subject to detect smaller disparities. Fig. 5 shows the thresholds for all of the subjects (excluding the seven who could not do the task reliably in any condition). 11 of the 16 subjects were able to detect the target circle at smaller disparities in the dynamic-lens session than in the fixed-lens session; the average thresholds were 1.8 and 2.5arcmin, respectively. The difference in thresholds was statistically reliable (Wilcoxon signed-rank test, p=0.028). The results indicate that minimizing the vergence-accommodation conflict by adjusting the power of lens in front of the eyes can aid the ability to detect small disparities. Thus, this aspect of visual performance is improved with the dynamic-lens technique.

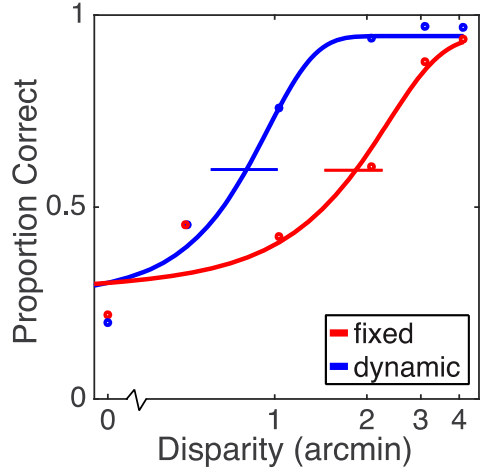

Fig. 4. Psychometric functions for subject 2 in the disparity-detection task with the dynamic-lens system. Proportion of correct responses is plotted as a function of the target circle's disparity. The red curve represents a best fit to the data from the fixed-lens condition. The blue curve represents a best fit to the data from the dynamic-lens condition. Error bars are 95% confidence intervals for the estimate of the disparity corresponding to 62.5% correct

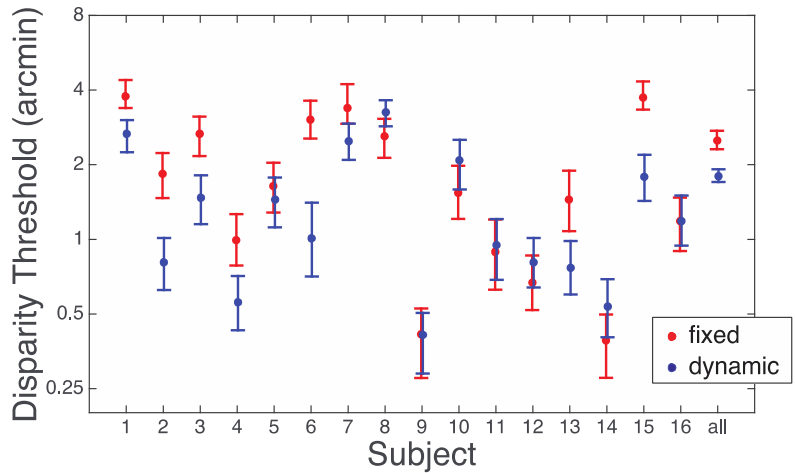

Fig. 5. Threshold disparity in the dynamic- and fixed-lens conditions. Thresholds are plotted for each subject as well as the threshold from the data pooled across subjects. Error bars represent 95% confident intervals.

The results from the symptom questionnaire are shown in the left half of Fig. 6. There were no systematic differences in reported symptoms between the dynamic- and fixed-lens sessions. The results from the comparison questionnaire were more revealing. There was a consistent preference for the dynamic-lens session ($p<0.05$, one-tailed Wilcoxon signed-rank test) and subjects reported relatively less fatigue ($p<0.05$), eye irritation ($p<0.05$), and headache ($p<0.05$) in that session. The observation of systematic differences in the comparison questionnaire, but not in the symptom questionnaire, is consistent with our previous experience [10,12,45]: Asking subjects to compare two experiences is more sensitive than asking subjects to rate one experience. In sum, the discomfort data clearly suggest that varying the power of lenses before the viewer's eyes, thereby reducing the vergence-accommodation conflict, can create a more comfortable viewing experience.

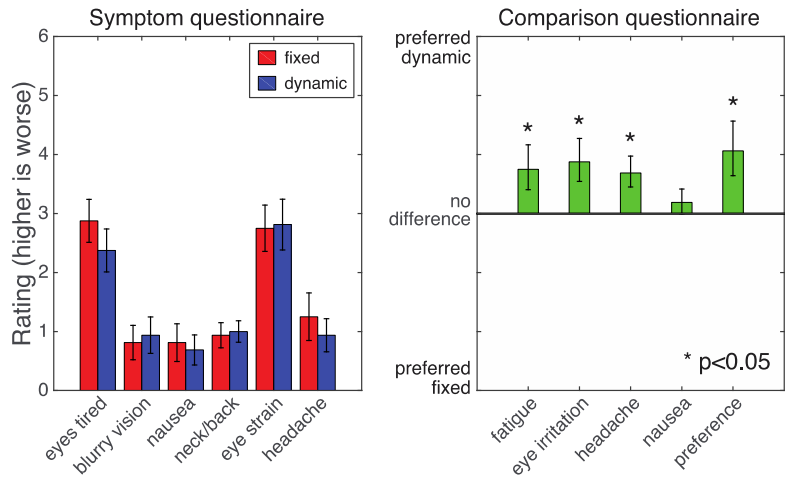

Fig. 6. Discomfort results from the dynamic-lens study. The results from the symptom and comparison questionnaires are shown in the left and right panels, respectively. The symptom questionnaire used a 0-6 rating scale for each of six questions, larger numbers indicating more uncomfortable symptoms. The average ratings are for the fixed- and dynamic-lens conditions are represented by the red and blue bars, respectively. The comparison questionnaire also used a 0-6 scale where 0 meant a strong preference for the fixed-lens condition and 6 a strong preference for the dynamic-lens condition. A one-tailed Wilcoxon signed-rank test was used to assess the statistical reliability of the differences in the comparison questionnaire data. * indicates p<0.05.

*3.3 Visual discomfort results for the monovision system*

We assessed subjects' ability to detect the target circle (the one with added disparity) among the circles in the approaching and receding diamond stimulus. We did so separately for the monovision session and the no-lens session. Fig. 7 shows one subject's psychometric data in the two sessions. This subject required a slightly larger disparity to perform the task reliably in the monovision session than in the no-lens session. Fig. 8 shows the disparity thresholds for each subject along with the average thresholds. Even though the differences were sometimes small, all eight subjects had lower thresholds in the no-lens condition. The differences were statistically reliable (two-tailed Wilcoxon signed-rank test, p=0.003). In the pooled data the thresholds were 0.90arcmin in the monovision session and 0.86arcmin in the no-lens session. We conclude that the ability to detect small disparities becomes slightly worse when the two eyes are optically corrected for different distances. This result is not surprising because previous work has shown that the blurring of one eye's image causes a reduction in stereoacuity [36,37]. Thus, the monovision technique does not seem to improve that aspect of visual performance. Indeed, it seems to make it slightly worse.

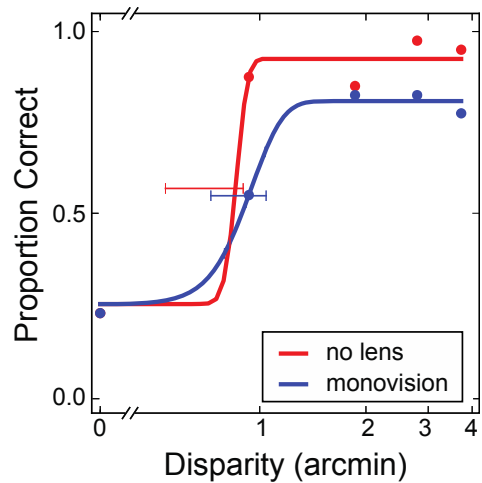

Fig. 7. Psychometric functions for one subject in the disparity-detection task in the monovision study. The data points indicate the proportion of correct responses as a function of the disparity of the target circle. The red curve represents the best-fitting function for the data in the no-lens condition. The blue curve represents the best function for the data in the monovision condition. Error bars are 95% confidence intervals for the estimate of the disparity corresponding to 62.5% correct.

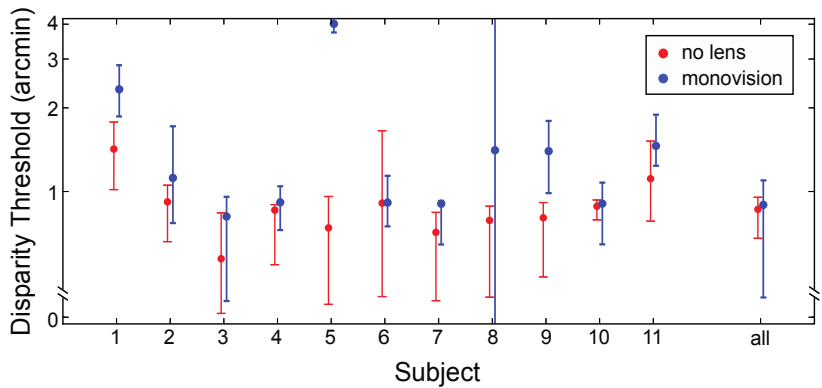

Fig. 8. Threshold disparities in the mononvision study. Red and blue symbols represent the average threshold data in the no-lens and monovision conditions, respectively. Thresholds are plotted for each subject as well as the average threshold obtained from pooling the data across subjects. Error bars represent 95% confidence intervals.

The monovision-study results from the symptom questionnaire are shown in the left half of Fig. 9. There were no systematic differences in reported symptoms in the no-lens and monovision sessions (two-tailed Wilcoxon signed-rank test). The results from the comparison questionnaire are shown in the right half of that figure. There were three statistically reliable differences in reported preferences between the two sessions (fatigue, p = 0.012; eye irritation, p = 0.023; overall preference, p = 0.014); in each case, the preference was for the no-lens condition. Thus, the questionnaire data did not reveal a reduction in visual discomfort in the monovision condition relative to the no-lens condition which mimicks a conventional stereoscopic display. We conclude that monovision does not offer an improvement in the comfort of the viewing experience. If anything, it makes discomfort worse.

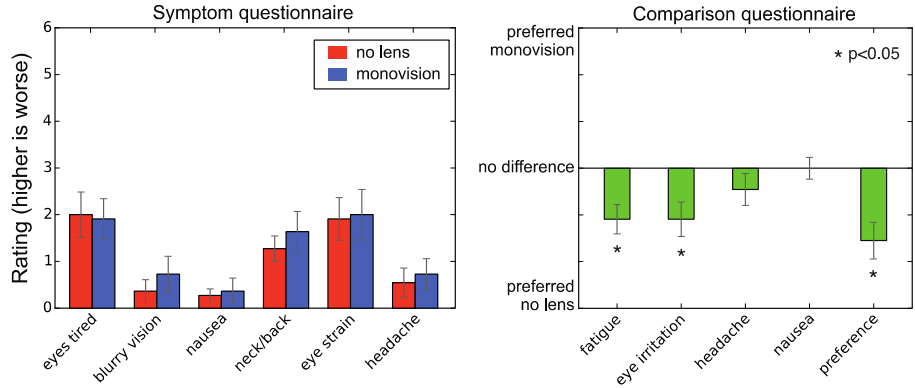

Fig. 9. Discomfort results for the monovision study. The results from the symptom and comparison questionnaires are shown in the left and right panels, respectively. The symptom questionnaire used a 0-6 scale for each of six questions; larger numbers indicate more uncomfortable symptoms. The average ratings are for the no-lens and monovision conditions are represented by the red and blue bars, respectively. The comparison questionnaire also used a 0-6 scale where 0 meant a strong preference for the monovision condition and 6 meant a strong preference for the no-lens condition. A two-tailed Wilcoxon signed-rank test was used to assess the statistical reliability of the differences in the comparison questionnaire data. * indicates differences for which $p < 0.05$.

## 4. Time-to-fuse experiment

### 4.1 Experimental Details

We also examined how the two display systems affect the ability to fuse binocular stimuli quickly. The stimuli were random-dot stereograms that contained a corrugation in depth that was oriented up and to the left (+20°) or up and to the right (-20°), like the one in Fig. 10. Subjects indicated the orientation after each stimulus presentation: a two-alternative, forced-choice task. Feedback was provided after each response. Fig. 11 is a schematic of the experimental procedure. The duration of the stimulus was varied based on the correctness of the subject's responses using an adaptive 1-up, 2-down staircase. Cumulative Gaussian functions were fit to the resulting data using a maximum-likelihood criterion [42]. From the fitted functions, we determined the duration required for 75% correct responding. The experiment lasted ~45 minutes including training and debriefing.

In the dynamic-lens experiment, the display was positioned 1.77m from the subject. The stimulus subtended 2.2°. The stimuli were generated in MATLAB and loaded into Python using the PsychoPy library. On each trial in the dynamic-lens experiment, a Maltese cross with zero disparity was first presented for 2sec. The random-dot stereogram stimulus then appeared at one of five distances relative to the screen: –0.25, +0.25, +0.75, or +1.5D. These disparity-specified distances relative to the subject were 0.31, 0.81, 1.31, or 2.06D (respectively 3.17, 1.23, 0.76, or 0.48m).

For the monovision experiment, the display was 0.5m (2D) from the subject. There was first a brief training phase that involved correctly identifying the orientation of three stimuli at increasing disparities, up to and including the disparities used in the experiment. Upon successful completion of this training phase, participants began the main experiment, in which the stereogram stimuli all appeared 1D in front of the display screen. The stimulus subtended 15.9°.

We excluded subjects from further analysis if they could not do better than 70% correct at the longest presentation times. This criterion led to the exclusion of 2 of 8 subjects in the dynamic-lens experiment and 5 of 18 in the monovision experiment.

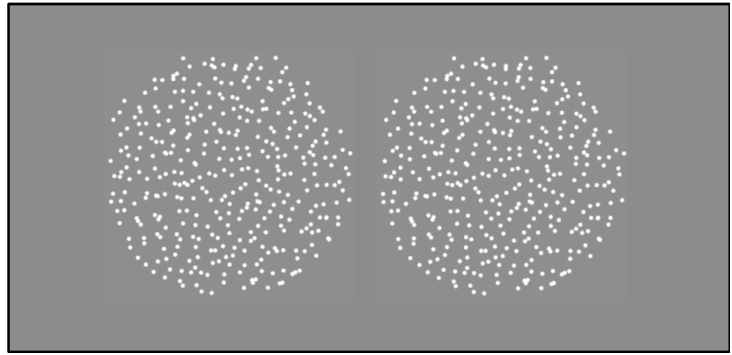

Fig. 10. Stimulus used in the time-to-fuse experiments. The stimulus subtended 2.2° or 15.9° in each eye. Cross-fuse to see a sinusoidal depth corrugation. Subjects indicated the orientation of the corrugation (a 2-alternative, forced-choice task).

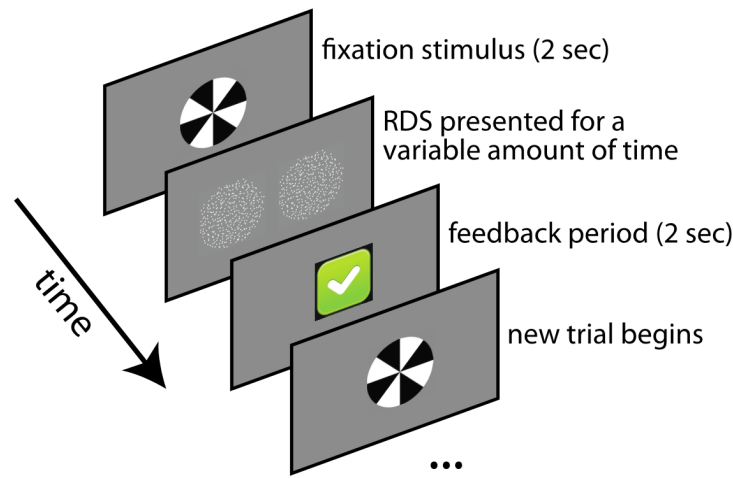

Fig. 11. Time-to-fuse experimental procedure. At the beginning of every trial, subjects fixated for 2sec on a Maltese cross. A random-dot stereogram then appeared at a random position in depth. Subjects were directed to fuse and indicate the orientation of the corrugations in the stimulus. In fixed-lens and no-lens trials, the focal distance was fixed at the screen. In dynamic-lens and monovision trials, the focal distance matched the distance of the stereogram. Dynamic- and fixed-lens trials were randomly interleaved throughout the experiment.

*4.2 Time-to-fuse results for dynamic-lens display*

For each disparity, we found the presentation time that was just necessary to fuse a binocular stimulus. Fig. 12 shows the psychometric data for a representative subject. This subject was able to achieve 75% correct performance with shorter stimulus durations in the dynamic-lens condition than in the fixed-lens condition. Fig. 13 shows individuals' threshold presentation times as a function of disparity for the dynamic-lens and fixed-lens conditions. With small disparities (–0.25, +0.25, and +0.75D), there was no significant difference between the dynamic and fixed conditions. With the largest disparity (+1.5D), however, there was a clear difference. The presentation time needed to fuse the stimulus was clearly greater in the fixed-lens condition (3.2sec) than in the dynamic condition (0.96sec). This finding is consistent with earlier observations that vergence and accommodation responses occur more quickly when the vergence and accommodation stimuli are consistent with one another [3,5]. We conclude that the proposed dynamic-lens system enables faster binocular fusion than conventional stereoscopic displays. Thus, this aspect of visual performance is improved by this system.

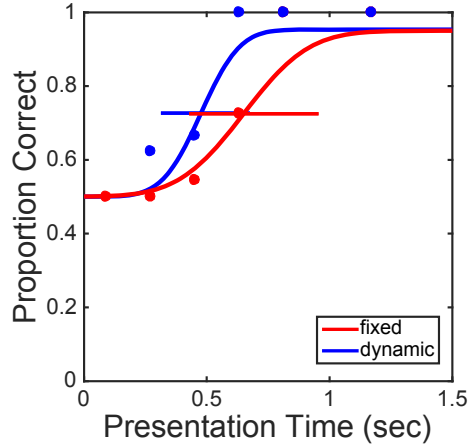

Fig. 12. Psychometric data for subject 3 in the time-to-fuse experiment with the dynamic-lens setup and a disparity of 1.5D. Proportion correct is plotted as a function of stimulus duration. Red represents the fixed-lens condition and blue the dynamic-lens condition. Error bars represent 95% confidence intervals on the duration required to reach 75% correct responding.

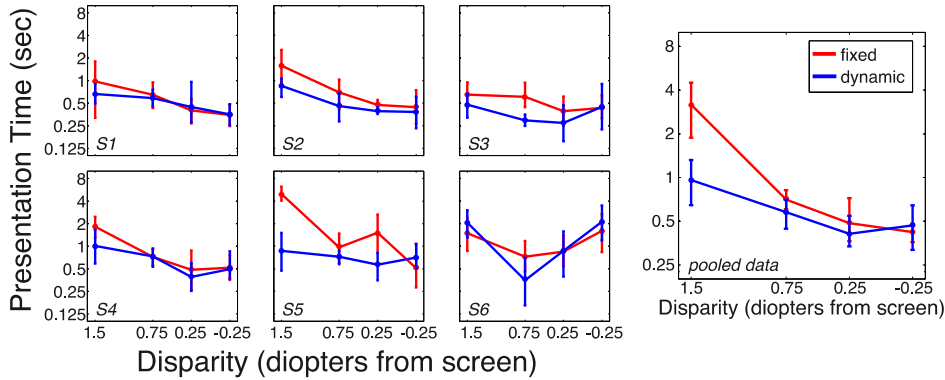

Fig. 13. Presentation times required to fuse in the dynamic-lens experiment. Each panel plots the stimulus duration required to achieve 75% correct responding as a function of the disparity of the stimulus relative to the screen. The six panels on the left show the data for individual subjects. The panel on the right shows the data once pooled across subjects. Error bars represent 95% confidence intervals.

*4.3 Time-to-fuse results for monovision display*

We again found the presentation time that was just needed to fuse the binocular stimulus, but this time in the monovision setup. Fig. 14 shows the psychometric data for a representative subject. This subject was no better at fusing the stimulus quickly in the monovision condition than in the no-lens condition. Fig. 15 shows individuals' threshold presentation times in the two conditions as well as the thresholds once pooled across subjects. Seven of the 14 subjects were able to fuse more quickly in the monovision condition than in the no-lens condition. The required presentation times were 0.26 and 0.31 sec in the pooled data for the monovision and no-lens conditions, respectively: a statistically insignificant difference. We conclude that the proposed monovision system does not enable faster binocular fusion than conventional stereoscopic displays. Thus, this aspect of visual performance is not improved by the monovision setup.

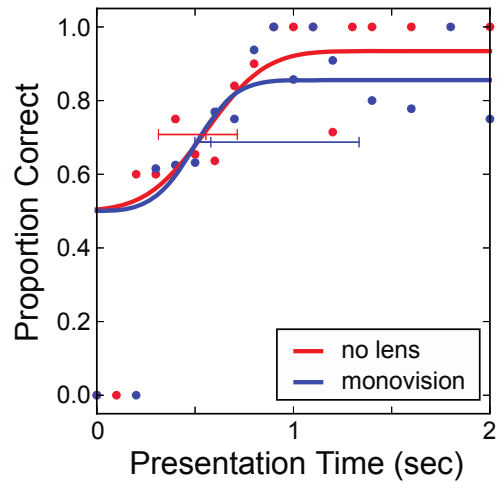

Fig. 14. Time-to-fuse data for subject 4 in the monovision experiment. Proportion correct responses are plotted as function of stimulus presentation time. Red represents the data in the no-lens condition and blue the data in the monovision condition. Error bars represent 95% confidence limits on the estimate of the presentation time at 75% correct.

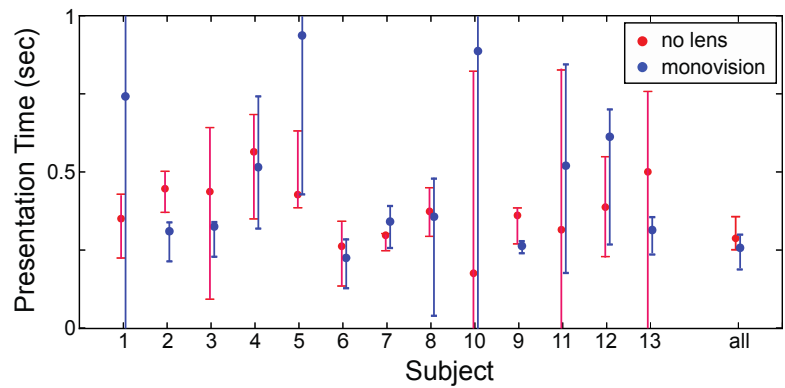

Fig. 15. Time-to-fuse thresholds for all subjects in the monovision experiment. The red and blue symbols represent the thresholds for the no-lens and monovision conditions, respectively. The thresholds from the data pooled across subjects are on the right. Error bars represent 95% confidence intervals.

## 5. Discussion

Our results show that visual discomfort can be reduced and visual performance can be improved by using the proposed dynamic-lens system. In this system the focal distance between the display and viewer's eyes is adjusted so that the stimulus to accommodation is similar to the stimulus to vergence. Our results also show that discomfort is not reduced and performance is not improved by using the proposed monovision system. In this system, the focal distance of one eye is made to differ from the other eye by 1D. We next discuss some implications of these results.

*5.1 Gaze prediction in the dynamic-lens system*

The dynamic-lens system relies on setting the power of the lenses in front of the viewer's eyes to make the accommodative response similar to the vergence response. But the vergence response depends on where in the simulated scene the viewer is looking, so one cannot know how to set the power of the variable lens without knowing the distance of the viewer's current fixation. We side-stepped this requirement in the experiments reported here by giving viewers a fixation point that moved in depth. Assuming they fixated accurately, we therefore always knew the distance of fixation and to what value to set the lens power. In a practical system the viewer would be allowed to fixate freely, so the system would have to measure or predict the viewer's fixation distance from moment to moment.

The obvious way to determine fixation distance over time is to measure it. Various eye-tracking systems have been developed, and have been incorporated into desk- and head-mounted devices [46,47]. For example, the Eyelink II eye tracker (SR Research, Ottawa, Ontario, Canada) can acquire binocular fixation data at a rate of 500Hz with an average error of 0.5° and a resolution of 0.01° [48-50]. The system finds the intersection of the left- and right-eye lines of sight to determine the distance of the fixated point. The estimation problem is complicated by the fact that the estimated lines of sight usually do not intersect so one has to estimate the true location by finding the position in space where the lines come closest to one another. One can in principle also estimate fixation distance by measuring the direction of the line of sight of one eye and then using knowledge of the simulated scene to determine the distance at which the line of sight intersects the scene.

Another way to estimate fixation distance is to use the contents of the simulated scene to predict what part will be fixated. Many saliency algorithms have been developed for this purpose. The algorithms use low-level features of the content to estimate salient objects in the scene, producing saliency map that indicates where the viewer is likely to look. For example, Itti and colleagues generated a biologically plausible model of saliency estimation by simulating "center-surround" operations similar to visual receptive fields [51]. Perazzi and co-workers demonstrated that two global measures of contrast—uniqueness of colors and spatial distribution of elements—could be used to estimate saliency [52]. Other methods use motion [53,54] or particular spatiotemporal features [55-57]. Disparity is also very useful for estimating saliency in stereoscopic displays [58].

Neither eye tracking nor saliency estimation would make accurate estimates of fixation distance all the time. So it is interesting to consider how often they would have to estimate distance correctly for the dynamic-lens system to help with discomfort and performance. We conducted simulations that show that vergence-accommodation conflict is on average reduced if the estimates are accurate to within a half diopter roughly 1/3 of the time (of course, the exact value depends on how big the depth volume is, how far away the screen is, the viewer's pupil diameter, how much the depth of the simulated scene varies, and so forth). Thus, eye tracking and/or saliency estimation may yield a practical system that reduces discomfort and increases visual performance relative to a conventional stereoscopic display.

*5.2 Depth-of-field simulation*

Depth-of-field blur could be easily integrated into the dynamic-lens system. Specifically, the the simulated scene could be rendered sharply at the distance of fixation and with increasing blur for parts of the scene that are progressively nearer and farther than the fixation distance. Adding depth-of-field blur to stereoscopic images increases relative visual comfort when viewers are trained to look at specific static or moving targets [59]. One team of researchers incorporated an eye tracker so that the depth-of-field rendering could be centered on the measured fixation distance [60]. Nonfixated areas were then blurred in a depth-dependent way. The results suggested that perceived visual quality was significantly improved in virtual scenes, but not in photographed scenes. Other researchers have demonstrated that tracking plus depth-of-field rendering can reduce visual discomfort under some conditions [61,62]. Thus, adding depth-of-field blur to our system, using knowledge of the viewer's fixation distance, could further reduce visual discomfort. Of course, inaccurate measurements of fixation distance could lead to adding blur to fixated objects, which could well be annoying and confusing [63].

**6. Conclusion**

The results of our study have shown that the dynamic-lens display may provide a viable means of reducing the vergence-accommodation conflict. Participants typically expressed a preference for this system, and exhibited a reduction in symptoms associated with sustained 3D viewing. They also performed binocular fusion tasks faster, and at smaller disparities.

The monovision system is a less complex means of presenting the observer with multiple focal distances, but this method did not significantly reduce viewer discomfort. This failure to reduce discomfort may have been caused by increased binocular rivalry due to inter-ocular differences in image sharpness.

**Acknowledgement**

We acknowledge funding from the EPSRC and the NIH. The data presented in this paper are available from **http://dx.doi.org/10.15128/736664863**.